\documentclass[letterpaper, 10 pt, conference]{ieeeconf}  % Comment this line out if you need a4paper

\IEEEoverridecommandlockouts                              % This command is only needed if 
\usepackage{graphics} % for pdf, bitmapped graphics files
\usepackage{epsfig} % for postscript graphics files
\usepackage{amsmath} % assumes amsmath package installed
\usepackage{amssymb}  % assumes amsmath package installed

\usepackage{hyperref}
\usepackage{xcolor}
\usepackage{graphicx}
\usepackage{subcaption}
\usepackage{ntheorem}
\newtheorem*{remark}{Remark}
\newtheorem{definition}{Definition}
\newtheorem{lemma}{Lemma}

\usepackage{mathbbol}

\usepackage{cancel}

\usepackage{etoolbox}

\makeatletter
\let\nobreakitem\item
\let\@nobreakitem\@item
\patchcmd{\nobreakitem}{\@item}{\@nobreakitem}{}{}
\patchcmd{\nobreakitem}{\@item}{\@nobreakitem}{}{}
\patchcmd{\@nobreakitem}{\@itempenalty}{\@M}{}{}
\patchcmd{\@xthm}{\ignorespaces}{\nobreak\ignorespaces}{}{}
\patchcmd{\@ythm}{\ignorespaces}{\nobreak\ignorespaces}{}{}

\renewtheoremstyle{break}%
  {\item{\theorem@headerfont
          ##1\
          ##2\theorem@separator}\hskip\labelsep\relax\nobreakitem}%
  {\item{\theorem@headerfont
          ##1\ ##2\ (##3)\theorem@separator}}%\hskip\labelsep\relax\nobreakitem}
\makeatother

\theoremheaderfont{\kern-0cm\normalfont\bfseries} 
\theoremstyle{break}
\newtheorem{theorem}{Theorem}

\usepackage{cite}
\makeatletter 
\pretocmd\@bibitem{\color{black}\csname keycolor#1\endcsname}{}{\fail}
\newcommand\citecolor[1]{\@namedef{keycolor#1}{}}
\makeatother
\title{\LARGE \bf
High-dimensional continuification control of large-scale multi-agent systems under limited sensing and perturbations}

\author{Gian Carlo Maffettone$^{1}$, Mario di Bernardo$^{1, 2, \dagger, *
}$, Maurizio Porfiri$^{3, \dagger, *}$% <-this % stops a space
\thanks{This work was developed with the economic support of MUR (Italian Ministry of University and Research) performing
the activities of the project PRIN 2022 “Machine-learning based control of complex multi-agent systems for search and rescue operations in natural
disasters (MENTOR) and of the National Science Foundation under Grant CMMI-1932187.}
\thanks{$^{1}$Scuola Superiore Meridionale, Naples, Italy.}
\thanks{$^{2}$Department of Electrical Engineering and Information Technology, University of Naples Federico II, Naples, Italy.}
\thanks{$^{3}$Center for Urban Science and Progress, Department of Biomedical engineering, Department of Mechanical and Aerospace Engineering, Tandon School of Engineering, New York University, New York, USA,}
\thanks{$^{\dagger}$These authors contributed equally.}
\thanks{$^{*}$For correspondance (mario.dibernado@unina.it, mporfiri@nyu.edu)}%
}
\usepackage[T1]{fontenc}

\begin{document}

\maketitle
\thispagestyle{empty}
\pagestyle{empty}

%%%%%%%%%%%%%%%%%%%%%%%%%%%%%%%%%%%%%%%%%%%%%%%%%%%%%%%%%%%%%%%%%%%%%%%%%%%%%%%%
\begin{abstract}
This paper investigates the robustness of a novel high-dimensional continuification control method for complex multi-agent systems. We begin by formulating a partial differential equation describing the spatio-temporal density dynamics of swarming agents. A stable control action for the density is then derived and validated under nominal conditions. Subsequently, we discretize this macroscopic strategy into actionable velocity inputs for the system's agents. Our analysis demonstrates the robustness of the approach beyond idealized assumptions of unlimited sensing and absence of perturbations.
\end{abstract}

%%%%%%%%%%%%%%%%%%%%%%%%%%%%%%%%%%%%%%%%%%%%%%%%%%%%%%%%%%%%%
%%%%%%%%%%%%%%%%%%%%%%%%%%%%%%%%%%%%%%%%%%%%%%%%%%%%%%%%%%%%%

\section{Introduction}\label{sec:introl}
\begin{figure}
     \centering
     \includegraphics[width=0.5\textwidth]{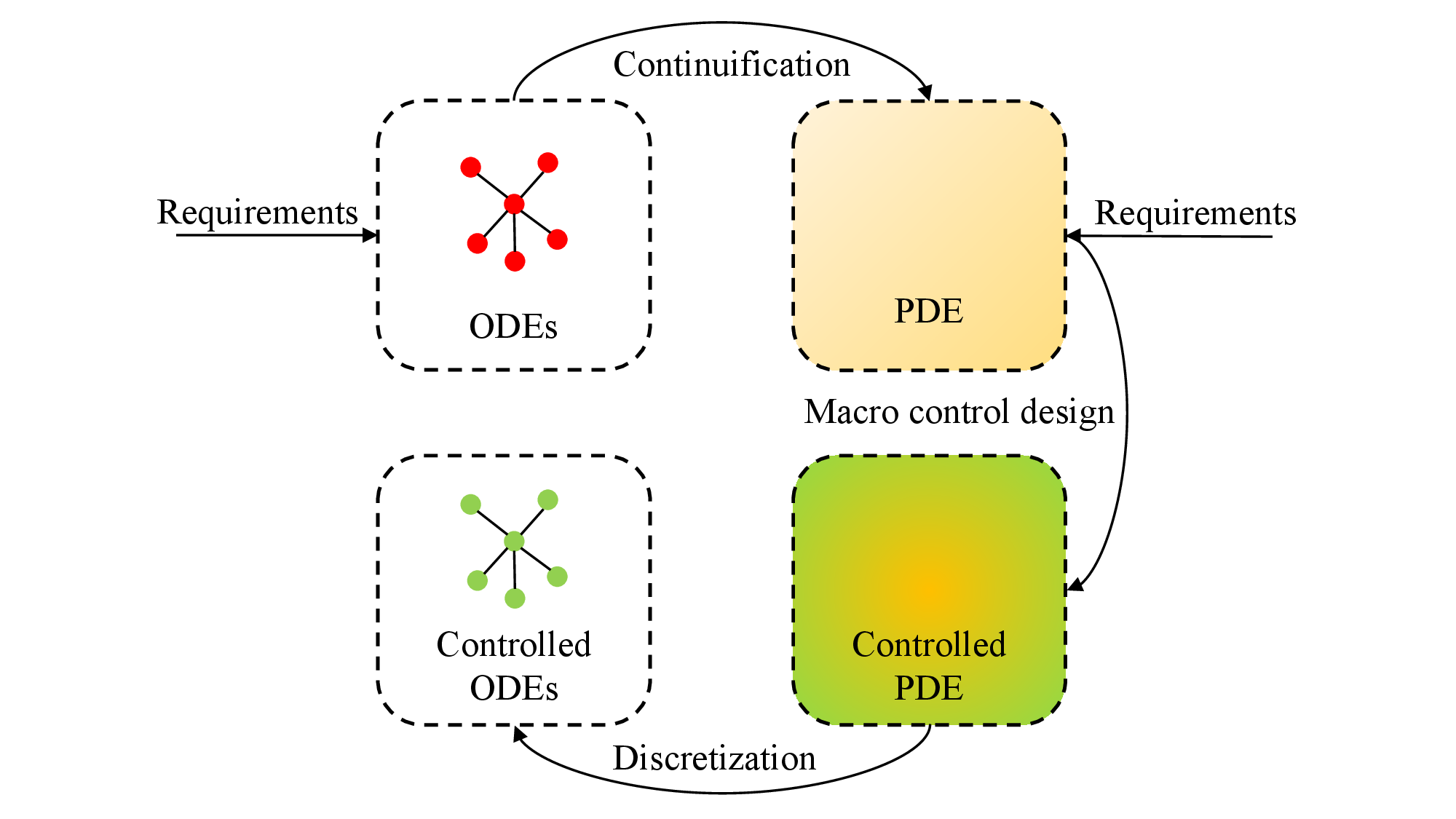}
     \caption{Continuification control scheme (inspired by \cite{nikitin2021continuation}).} 
     \label{fig:continuification}
\end{figure}

The continuification-based control approach generates discrete microscopic control protocols through the continuum, macroscopic approximation of large-scale multi-agent systems \cite{maffettone2022continuification, nikitin2021continuation}. Within such an approach, the system dynamics is first described using a traditional agent-based framework, in the form of a large set of ordinary differential equations (ODEs). Next, a macroscopic description of the emergent behavior is derived, often as a small set of partial differential equations (PDEs), under the assumption of an infinite number of agents. Although the control design is developed based on this macroscopic approximation of the system behavior, the macroscopic control action is ultimately discretized into actionable microscopic inputs, as illustrated in Fig. \ref{fig:continuification}. 

One of the most constraining assumptions of this approach is the premise that agents possess unlimited sensing capabilities, enabling them to influence one another even over long distances. Also, the potential existence of perturbations or disturbances that affect the system dynamics is frequently overlooked.
In this paper, we expand upon the analysis conducted in \cite{maffettone2023continuification}, which focused on one-dimensional domains, to evaluate the robustness of the high-dimensional continuification control strategy elucidated in \cite{carlo2023mixed}. Our aim is to relax ($i$) the assumption of agents having unlimited access to knowledge on the other agents' dynamics and ($ii$) the assumption of absence of perturbations. Specifically, by considering appropriate conditions and utilizing the macroscopic system formulation, we establish analytical guarantees of semi-global and bounded stability under limited sensing capabilities and perturbations. Each scenario is rigorously validated through comprehensive simulations.

The rest of the paper is organized as follows. In Section \ref{sec:math_prel}, we provide useful mathematical notation; in Sections \ref{sec:the_model}, \ref{subsec:probstatement} and \ref{sec:methods}, we briefly recall the theoretical framework that is discussed in \cite{carlo2023mixed}; in Sections \ref{sec:lim_sens} and \ref{sec:pert}, we study the robustness of the solution.
%with respect to limited sensing and perturbations, respectively. 
Analytical results are numerically validated in Section \ref{sec:num_valid}.

\section{Mathematical preliminaries}\label{sec:math_prel}
Here, we give some mathematical notation that will be used throughout the paper.
We define $\Omega:=[-\pi, \pi]^d$, with $d=1,2,3$ the periodic cube of side $2\pi$. The case $d=1$ corresponds to the unit circle, $d=2$ to the periodic square, and $d=3$ to the periodic cube. We refer to $\partial\Omega$ for indicating $\Omega$'s boundary.
\begin{definition}[$L^p$-norm on $\Omega$ \cite{Axler2020}]\label{def:Lp_norm}
    Given a scalar function of $h:\Omega\times\mathbb{R}_{\geq0}\rightarrow\mathbb{R}$, we define its $L^p$-norm as
\begin{align}
    \Vert h(\cdot, t)\Vert_p := \left(\int_{\Omega} \vert h(\mathbf{x}, t)\vert^p \,\mathrm{d}\mathbf{x}\right)^{1/p}.
\end{align}
The case $p=\infty$ is defined as
\begin{align}
    \Vert h(\cdot, t)\Vert_\infty :=\mathrm{ess} \,\mathrm{sup}_\Omega \vert h(\mathbf{x}, t)\vert.
\end{align}
For brevity, we also denote these norms as $\Vert h\Vert_p$, without explicitly indicating their dependencies. 
\end{definition}
Notice that a vector-valued function is said to be $L^p$-bounded if all its components have a bounded $L^p$-norm.

\begin{lemma} [Holder's inequality \cite{Axler2020}]\label{th:holder}
    Given $n$ $L^p$ functions, $f_i$, with $i=1, 2, \dots n$, we have
    \begin{align}
        \left\Vert \prod_{i=1}^n f_i\right\Vert_1 \leq \prod_{i=1}^n \Vert f_i\Vert_{p_i}, \;\;\mathrm{if } \;\; \sum_{i=1}^n \frac{1}{p_i} = 1.
    \end{align}
    For instance, if $n=2$, $\Vert f_1f_2\Vert_1\leq \Vert f_1\Vert_2 \Vert f_2\Vert_2$ or $\Vert f_1f_2\Vert_1\leq \Vert f_1\Vert_1 \Vert f_2\Vert_\infty$.
\end{lemma}
\begin{lemma}[Minkowsky inequality \cite{Axler2020}]\label{th:Minkowsky}
    Given two $L^p$ functions, $f$ and $g$, the following inequality holds:
    \begin{align}
        \Vert f + g \Vert_p \leq \Vert f\Vert_p + \Vert g\Vert_p,
    \end{align}
    for $1\leq p\leq \infty$. 
    %This proves $L^p$ spaces are normed vector spaces, as it generalize the triangular inequality to $L^p$ functions.
\end{lemma}
% \begin{lemma}\label{lemma:lem1}
% Given Theorem \ref{th:Minkowsky}, also the following inequality holds
% \begin{align}
%     \Vert f - g \Vert_p \leq \Vert f\Vert_p + \Vert g\Vert_p,
% \end{align}
% for $1\leq p\leq \infty$.
% \end{lemma}
We denote with subscripts $t$ and $x$ time and space partial derivatives. We indicate gradient as $\nabla (\cdot)$, divergence as $\nabla \cdot (\cdot)$, curl as $\nabla \times (\cdot)$, and Laplacian as $\nabla^2 (\cdot)$.
%given avector (scalar) valued function $\mathbf{h}$ ($h$), we denote its Jacobian (gradient) as $\nabla \mathbf{h}$ ($\nabla h$), its divergence as $\nabla \cdot \mathbf{h}$, its curl as $\nabla \times \mathbf{h}$, and its Laplacian as $\nabla^2\mathbf{h}$.

We denote with \textquotedblleft{} $*$ \textquotedblright{} the convolution operator. When referring to periodic domains and functions, the operator needs to be interpreted as a circular convolution \cite{jeruchim2006simulation}. We remark that the circular convolution is itself periodic. It can be shown \cite{jeruchim2006simulation} that we have
\begin{align}
    (f*g)_x(x)  = (f_x*g)(x) = (f*g_x)(x).
\end{align}

\begin{lemma} [Young's convolution inequality \cite{Axler2020}] \label{th:young_inequality}
    Given two functions, $f\in L^p$ and $g \in L^q$, we have
    \begin{align}
        \Vert f * g\Vert_r \leq \Vert f\Vert_p \,\Vert g\Vert_q,\;\;\mathrm{if }\;\;\frac{1}{p} + \frac{1}{q} = \frac{1}{r} + 1,
    \end{align}
    where $1\leq p,q,r \leq\infty$. %For instance, it holds that $\Vert f*g \Vert_\infty \leq \Vert f \Vert_2\Vert g \Vert_2$. 
\end{lemma}

%\begin{definition}\label{lem:young_deriv}
% It can be proved \cite{jeruchim2006simulation} that    given the convolution of two functions, say $(f*g)(x)$, we can compute its derivative as
% \begin{align}
%     (f*g)_x(x)  = (f_x*g)(x) = (f*g_x)(x).
% \end{align}
\begin{lemma}[Comparison lemma \cite{khalil2002nonlinear}]\label{lemma:comparison_lemma}
    Given  a scalar ODE $v_t = f(t, v)$, with $v(t_0) =
    v_0$, where $f$ is continuous in $t$ and locally Lipschitz in $v$, if a scalar function $u(t)$ fulfills the differential inequality
    \begin{align}
        u_t \leq f(t, u(t)), \;\;u(t_0)\leq v_0,
    \end{align}
    then,
    \begin{align}
        u(t)\leq v(t),\;\;\forall\,t\geq t_0.
    \end{align}
\end{lemma}
\begin{lemma}[Chapter 1.2 of \cite{griffiths2023introduction}]\label{lem:div_rel}
    Given a scalar function $\psi$, and a vector field $\mathbf{A}$, the following identity holds:
    \begin{align}
        \nabla \cdot (\psi \mathbf{A}) = \psi \nabla \cdot \mathbf{A} + \nabla\psi \cdot \mathbf{A}.
    \end{align}
    %where we used the nabla operator to indicate gradients and divergence.
\end{lemma}
\begin{lemma}\label{lem:surf_int}
    For any function $h$ that is periodic on $\partial \Omega$, we have
    \begin{align}
        \int_{\partial \Omega} h(\mathbf{x}) \cdot \mathbf{\hat{n}}\,\mathrm{d}\mathbf{x} = 0,
    \end{align}
    where $\hat{\mathbf{n}}$ is the is the outward pointing unit normal vector at each point on the boundary (by decomposing the integral on each side of the domain with the appropriate sign).
\end{lemma}
We denote by $\mathbb{n} = (n_1,\dots, n_d)$ the $d$-dimensional multi-index, consisting in the tuple of dimension $d$, with $n_i\in\mathbb{Z}$. Thus, $\mathbf{n} = [n_1, \dots, n_d]$ is the row vector associated with $\mathbb{n}$.

\section{Model and Problem Statement}\label{sec:the_model}
We consider $N$ dynamical units moving in $\Omega$.
The agents' dynamics are modeled using the \textit{kinematic assumption} \cite{viscek1995, bernoff2011primer} (i.e., neglecting acceleration and considering a drag force proportional to the velocity). Specifically, we set
\begin{align}\label{eq:d_dimensional_micro}
    \dot{\mathbf{x}}_i = \sum_{k=1}^N \mathbf{f}\left(\{\mathbf{x}_{i}, \mathbf{x}_{k}\}\right) + \mathbf{u}_i, \;\;\;  i=1,\dots,N,
\end{align}
% \begin{figure}
%     \centering
%     \includegraphics[width=0.5\textwidth]{figure1.eps}
%     \caption{Continuification control scheme (inspired by \cite{nikitin2021continuation}). The schemes describes all the stages of the solution: ($i$) continuification, ($ii$) macroscopic control design, and ($iii$) discretization. } 
%     \label{fig:continuification}
% \end{figure}
where $\mathbf{x}_i \in \Omega$ is the $i$-th agent's position, and $\{\mathbf{x}_{i}, \mathbf{x}_{k}\}$ is the relative { position} between agent $i$ and {$k$}, wrapped to have values in $\Omega$ {($d$-dimensional extension of what was defined in  \cite{maffettone2022continuification})},
%Since the domain is periodic, the relative distance between two locations is not uniquely defined. Here we consider $\{\mathbf{x}_{i}, \mathbf{x}_{j}\}$ to be the vector with the smallest norm pointing from $\mathbf{x}_j$ to  $\mathbf{x}_i$. Such a relative distance takes values in $[-\pi, \pi]^d$. 
$\mathbf{f}:\Omega\rightarrow \mathbb{R}^d$ is a periodic velocity interaction kernel modeling pairwise interactions between the agents (repulsion, attraction, or a mix of the two at different ranges -- see \cite{carlo2023mixed} for more details),  and $\mathbf{u}_i$ is a velocity control input designed so as to fulfill some control problem. 
% Furthermore, we assume $\mathbf{f}(\mathbf{z}) = -\nabla F(\mathbf{z})$, where $F: \mathbb{R}^d \rightarrow \mathbb{R}$ is a
% \textit{soft-core} potential, meaning that $\mathbf{f}(\mathbf{0}) = \mathbf{0}$. The Morse potential
%, vastly used in the literature \cite{bernoff2011primer, Dorsogna2006}, 
%is a choice of this kind.
%Two examples are shown in Fig. \ref{fig::2Dkernels} for $d=2$. A repulsive kernel in Fig. \ref{subfig::2d_kern_rep}, and a Morse kernel, with long-range attraction and short-range repulsion, in Fig. \ref{subfig::2d_kern_stab_Agg}. 
%We remark that, in the absence of control, agents subject to a repulsive kernel will spread in $\Omega$ until reaching an equilibrium configuration. Agents subject to a { Morse-like} kernel { (long-range attraction and short-range repulsion)}, will reach an aggregated compact formation (see \cite{bernoff2011primer} for a comprehensive description of the uncontrolled problem with 1D examples). 
% \begin{figure}
% \centering
% \subfloat[]{\includegraphics[width=0.24\textwidth]{figure4a.eps}%
% \label{subfig::2d_kern_rep}}
% \hfil
% \subfloat[]{\includegraphics[width=0.24\textwidth]{figure4b.eps}%
% \label{subfig::2d_kern_stab_Agg}}
% \caption{2D velocity interaction kernel: (a) repulsion, and (b) attraction at long range and repulsion at short range. Green regions represent repulsion, while red regions attraction. The length of the arrows is associated to the intensity of the interaction.}
% \label{fig::2Dkernels}
% \end{figure}

\subsection{Problem statement}\label{subsec:probstatement}
The problem is that of selecting a set of control inputs $\mathbf{u}_i$ allowing the agents to organize into a desired macroscopic configuration on $\Omega$. 
Specifically, given some desired periodic smooth density profile, $\rho^\text{d}(\mathbf{x}, t)$, associated with the target agents' configuration, the problem can be reformulated as that of finding a set of control inputs $\mathbf{u}_i,\ i=1,2,\dots,N$ in \eqref{eq:d_dimensional_micro} such that
\begin{equation}
    \lim_{t\rightarrow \infty} \Vert{\rho^\text{d}(\cdot, t)}-\rho(\cdot, t)\Vert_2=0,
\end{equation} 
for agents starting from any initial configuration.% $\mathbf{x}_i(0)=\mathbf{x}_{i0}, \ i=1,2,\ldots,N$.

\section{High-dimensional Continuification Control}\label{sec:methods}
In this section, we briefly recall the theoretical steps of the continuification-based control approach presented in \cite{carlo2023mixed}.
\subsection{Continuification}
We recast the microscopic dynamics of the agents \eqref{eq:d_dimensional_micro} as the mass balance equation \cite{bernoff2011primer, maffettone2022continuification}
\begin{align}\label{eq:d_dimensional_macro}
    \rho_t(\mathbf{x}, t) + \nabla \cdot \left[\rho(\mathbf{x}, t)\mathbf{V}(\mathbf{x}, t)\right] = q(\mathbf{x}, t), 
\end{align}
where
\begin{align}
    \mathbf{V}(\mathbf{x}, t) = \int_{\Omega} \mathbf{f}\left(\{\mathbf{x}, \mathbf{z}\}\right) \rho(\mathbf{z}, t) \,\mathrm{d}\mathbf{\mathbf{z}} = (\mathbf{f}*\rho)(\mathbf{x}, t).
\end{align}
represents the characteristic velocity field encapsulating the interactions between the agents in the continuum. 
%\textquotedblleft{} $*$ \textquotedblright{} is the convolution operator. As it is applied to periodic domains and functions, it needs to be interpreted as the circular convolution \cite{jeruchim2006simulation}. 
The scalar function $q$ is the macroscopic control action. %Although written as a mass source/sink for simplifying derivations, it will be in the end recast as a velocity field.

We require periodicity of $\rho$ on $\partial\Omega$ $\forall t \in \mathbb{R}_{\geq 0}$ and that $\rho(\mathbf{x}, 0) = \rho_0(\mathbf{x})$.
We remark that $\mathbf{V}$ is periodic by construction, as it comes from a circular convolution. Thus, for the periodicity of the density it is enough to ensure $\left(\int_\Omega \rho(\mathbf{x}, t) \,\mathrm{d}\mathbf{x}\right)_t = 0$, when $q = 0$ (using the divergence theorem and the periodicity of the flux).

\subsection{Macroscopic Control Design}
We assume the desired density profile, $\rho^\mathrm{d}(\mathbf{x}, t)$,  obeys to the mass conservation law
\begin{align}\label{eq:d_dimensional_ref_dyn}
    \rho^\mathrm{d}_t(\mathbf{x}, t) + \nabla \cdot \left[\rho^\mathrm{d}(\mathbf{x}, t)\mathbf{V}^\mathrm{d}(\mathbf{x}, t)\right] = 0,
\end{align}
where 
\begin{align}
    \mathbf{V}^\mathrm{d}(\mathbf{x}, t) = \int_{\Omega} \mathbf{f}\left(\{\mathbf{x, \mathbf{z}}\}\right) \rho^\mathrm{d}(\mathbf{z}, t) \,\mathrm{d}\mathbf{\mathbf{z}} = (\mathbf{f}*\rho^\mathrm{d})(\mathbf{x}, t).
\end{align}
Periodic boundary conditions and initial condition for \eqref{eq:d_dimensional_ref_dyn} are set similarly to those of \eqref{eq:d_dimensional_macro}.
Furthermore, we define the error function $e(\mathbf{x}, t) := \rho^\mathrm{d}(\mathbf{x}, t)-\rho(\mathbf{x}, t)$.

\begin{theorem}[Macroscopic convergence]\label{th_GAS_d_dim}
    Choosing 
    \begin{multline}\label{eq:d_dimensional_control}
        q(\mathbf{x}, t) = K_\mathrm{p} e(\mathbf{x}, t) - \nabla \cdot \left[e(\mathbf{x}, t) \mathbf{V}^\mathrm{d}(\mathbf{x}, t) \right] \\- \nabla \cdot \left[\rho(\mathbf{x}, t) \mathbf{V}^\mathrm{e}(\mathbf{x}, t) \right],
    \end{multline}
    where $K_\mathrm{p}$ is a positive control gain and $\mathbf{V}^\mathrm{e}(\mathbf{x}, t) = (\mathbf{f}*e)(\mathbf{x}, t) $, the error dynamics globally asymptotically converges to 0 
    \begin{align}\label{eq:d_dim_lim}
        \lim_{t\to\infty} e(\mathbf{x}, t) = 0 \;\;\; \forall \,e(\mathbf{x}, 0).
    \end{align}
\end{theorem}
\begin{proof}
    See Theorem 1 in\cite{carlo2023mixed}.
    % We can compute the error dynamics by subtracting \eqref{eq:d_dimensional_macro} from \eqref{eq:d_dimensional_ref_dyn}, resulting in
    % \begin{multline}\label{eq:err_dyn_1}
    %     e_t(\mathbf{x}, t) + \nabla \cdot \left[\rho^\mathrm{d}(\mathbf{x}, t) \mathbf{V}^\mathrm{d}(\mathbf{x}, t) \right] - \\\nabla \cdot \left[\rho(\mathbf{x}, t) \mathbf{V}(\mathbf{x}, t) \right] = -q(\mathbf{x}, t).
    % \end{multline}
    % The error function $e(\mathbf{x}, t)$ is periodic on $\partial \Omega$ $\forall t\in\mathbb{R}_{\geq0}$ and $e(\mathbf{x}, 0) = \rho^\mathrm{d}(\mathbf{x}, 0)-\rho(\mathbf{x}, 0)$. Then, taking into account that $\rho = \rho^\mathrm{d}-e$, and $\mathbf{V} = \mathbf{V}^\mathrm{d}-\mathbf{V}^\mathrm{e}$, we rewrite \eqref{eq:err_dyn_1} as
    % \begin{multline}
    %     e_t(\mathbf{x}, t) + \nabla \cdot \left[e(\mathbf{x}, t) \mathbf{V}^\mathrm{d}(\mathbf{x}, t) \right] + \\\nabla \cdot \left[\rho(\mathbf{x}, t) \mathbf{V}^\mathrm{e}(\mathbf{x}, t) \right]= -q(\mathbf{x}, t).
    % \end{multline}
    % Plugging in \eqref{eq:d_dimensional_control}, we get
    % \begin{align}
    %     e_t(\mathbf{x}, t) = -K_\mathrm{p}e(\mathbf{x}, t).
    % \end{align}
    % Since  $K_\mathrm{p}>0$, \eqref{eq:d_dim_lim} holds.
\end{proof}

\subsection{Discretization and Microscopic control}
In order to dicretize the macroscopic control action $q$, we first recast the macroscopic controlled model as
\begin{align}\label{eq:controlled_by_U}
    \rho_t(\mathbf{x}, t) + \nabla \cdot \left[\rho(\mathbf{x}, t) \left(\mathbf{V}(\mathbf{x}, t) + \mathbf{U}(\mathbf{x}, t)\right)\right] = 0,
\end{align}
where $\mathbf{U}$ is a controlled velocity field, that incorporates the control action.
Equation \eqref{eq:controlled_by_U} is equivalent to \eqref{eq:d_dimensional_macro}, if
\begin{align}\label{eq:divergence_of_U}
    \nabla \cdot \left[\rho(\mathbf{x}, t)\mathbf{U}(\mathbf{x}, t)\right] = -q(\mathbf{x}, t).
\end{align}
In contrast to the case of $d = 1$ discussed in \cite{maffettone2022continuification}, equation \eqref{eq:divergence_of_U} is insufficient to uniquely determine $\mathbf{U}$ from $q$ since it represents only a scalar relationship. Hence, we define the flux $\mathbf{w}(\mathbf{x}, t):= \rho(\mathbf{x}, t)\mathbf{U}(\mathbf{x}, t)$, and close the problem by adding an extra differential constraint on the curl of $\mathbf{w}$. Namely, we consider the set of equations
\begin{align}\label{eq:w}
    \begin{cases}
        \nabla \cdot \mathbf{w}(\mathbf{x}, t) = -q(\mathbf{x}, t)\\
    \nabla \times \mathbf{w}(\mathbf{x}, t) = 0
    \end{cases}
\end{align}
For problem \eqref{eq:w} to be well-posed, we require $\mathbf{w}$ to be periodic on $\partial \Omega$. 
%Notice that \eqref{eq:w} is a purely spatial problem, as no time derivatives are involved.
We remark that the choice of closing the problem using the irrotationality condition is arbitrary, and other closures can be considered. 
%This specific one allows not to introduce vorticity into the velocity field we are looking for. 

Similarly to the typical approach used in electrostatic \cite{griffiths2023introduction}, since $\Omega$ is simply connected and $\nabla { \times} \mathbf{w} = 0$, we can express $\mathbf{w}$ using the scalar potential $\varphi$. Specifically, we pose $\mathbf{w} (\mathbf{x}, t) = -\nabla\varphi(\mathbf{x}, t)$. Plugging this into the divergence relation in \eqref{eq:w}, we recast \eqref{eq:w} as the Poisson equation
\begin{align}\label{eq:Poisson}
    \nabla^2 \varphi (\mathbf{x}, t) = q (\mathbf{x}, t).
\end{align}
Problem \eqref{eq:Poisson} is characterized by the periodicity of $\nabla \varphi(\mathbf{x}, t)$ on $\partial \Omega$. Thus, \eqref{eq:Poisson}, together with its boundary conditions, defines $\varphi$ up to a constant $C$. Since we are interested in computing $\mathbf{w} = -\nabla \varphi$, the value of $C$ is irrelevant.
We solve the Poisson problem \eqref{eq:Poisson} in $\Omega$ by expanding $\varphi$ as a Fourier series. Specifically, we get
\begin{align}\label{eq:phi}
    \varphi(\mathbf{x}) = \sum_{\mathbb{m}\in \mathbb{Z}^d} \gamma_\mathbb{m} \,\mathrm{e}^{j\mathbf{m}\cdot \mathbf{x}} + C,
\end{align}
where, $\gamma_\mathbf{m}$ is the $\mathbb{m}$-th Fourier coefficient, $j$ is the imaginary unit, and $\mathbf{x}$ is assumed to be a column. Given this expression for the potential, we write its Laplacian as
\begin{align}\label{eq:fourier_laplacian}
    \nabla^2\varphi(\mathbf{x}) = \sum_{\mathbb{m}\in \mathbb{Z}^d} \gamma_\mathbb{m} \Vert \mathbf{m}\Vert^2 \mathrm{e}^{j\mathbf{m}\cdot \mathbf{x}}.
\end{align}
Next, we can apply Fourier series to $q$, resulting in
\begin{align}\label{eq:fourier_q}
    q(\mathbf{x}) = \sum_{\mathbb{m}\in \mathbb{Z}^d} c_\mathbb{m} \,\mathrm{e}^{j\mathbf{m}\cdot \mathbf{x}}, 
\end{align}
where, since at time $t$ the function $q$ is known, we can also express the coefficients as
\begin{align}
    c_\mathbb{m} = \frac{1}{(2\pi)^d}\int_\Omega q(\mathbf{x}) \mathrm{e}^{-j\mathbf{m}\cdot \mathbf{x}}\,\mathrm{d}\mathbf{x}.
\end{align}
Then, recalling \eqref{eq:Poisson}, we express the coefficients of the Fourier series of the potential $\varphi$ as
\begin{align}
    \gamma_\mathbb{m} = -\frac{c_\mathbb{m}}{\Vert \mathbf{m}\Vert^2}.
\end{align}

For a practical algorithmic implementation, when computing $\varphi$, we approximate it truncating the summation after some large $M$. Then, $\mathbf{w} = -\nabla \varphi$ and, $\mathbf{U} = \mathbf{w}/\rho$. 
%Such derivations need to take place at each $t$. 
%For the implementation, 
% only considering the first $M$ (with $M$ sufficiently large) terms of the infinite summations in \eqref{eq:phi}. 

Finally, we  compute the microscopic control inputs for the discrete set of agents by spatially sampling  $\mathbf{U}(\mathbf{x}, t)$, that is
\begin{align}
    \mathbf{u}_i(t) = \mathbf{U}(\mathbf{x}_i, t), \;\;\;\; i=1,2,\dots,N.
\end{align}
%Note that our dicscretization procedure is different from the one that is proposed in \cite{nikitin2021continuation}.
% \begin{remark}
%     The macroscopic control action $q$ is based on non-local terms like $\mathbf{V}^\mathrm{d}$ and $\mathbf{V}^\mathrm{e}$, making the control action exerted at $\mathbf{x}$ depending on the error everywhere else in $\Omega$.
% \end{remark}
\begin{remark}
    The macroscopic velocity field $\mathbf{U}$ is well-defined only when $\rho \neq 0$. As $\mathbf{U}$ will be sampled at the agents locations, i.e. where the density is different from 0, we know $\mathbf{U}$ is well defined where it is needed. Moreover, for implementation, we will finally estimate the density starting from the agents positions with a Gaussian estimation kernel, making it always different from 0.
\end{remark}

\section{Robustness to limited sensing}\label{sec:lim_sens}
The macroscopic control law we propose in \eqref{eq:d_dimensional_control} is based on the non-local convolution term $\mathbf{V}^\mathrm{e}$. For computing such a control action, agents need to know $e$ everywhere in $\Omega$, meaning that they need to posses sensing capabilities to cover the whole set $\Omega$.

Here, we relax this unrealistic assumption, considering agents only possess a limited sensing radius $\Delta$, i.e., they can only measure $e$ in a neighborhood of radius $\Delta$ located about their positions. We model such a case by considering a modified interaction kernel defined as
\begin{align}
    \hat{\mathbf{f}}(\mathbf{z}) = \begin{cases}
        \mathbf{f}(\mathbf{z}) \;\;&\mathrm{if} \,\Vert \mathbf{z}\Vert_2\leq \Delta\\
        \mathbf{0} &\mathrm{otherwise}
    \end{cases}.
\end{align}
In this scenario, the macroscopic control law takes the form
\begin{multline}\label{eq:q_lim_sens}
        \hat{q}(\mathbf{x}, t) = K_\mathrm{p} e(\mathbf{x}, t) - \nabla \cdot \left[e(\mathbf{x}, t) \mathbf{V}^\mathrm{d}(\mathbf{x}, t) \right] \\- \nabla \cdot \left[\rho(\mathbf{x}, t) \hat{\mathbf{V}}^\mathrm{e}(\mathbf{x}, t) \right],
\end{multline}
where $\hat{\mathbf{V}}^\mathrm{e} = (\hat{\mathbf{f}}*e)$.
    Under control action \eqref{eq:q_lim_sens}, the error system dynamics may be written as
 \begin{multline}\label{eq:e_lim_sens}
    e_t(\mathbf{x}, t) = -K_\mathrm{p}e(\mathbf{x}, t) + \nabla \cdot \left[\rho^\mathrm{d}(\mathbf{x}, t)\Tilde{\mathbf{V}}(\mathbf{x}, t)\right]
    \\- \nabla \cdot \left[e(\mathbf{x}, t)\Tilde{\mathbf{V}}(\mathbf{x}, t)\right],
\end{multline}
where $\Tilde{\mathbf{V}} = (\mathbf{g}*e)$ and $\mathbf{g}=\mathbf{\hat{f}}-\mathbf{f}$.

Now, we provide some lemmas, that will be used for studying the stability properties of the perturbed error system \eqref{eq:e_lim_sens}.
\begin{lemma}\label{lem:lim_sens_lem1}
The following inequality holds:
    $$
    \Vert \nabla \cdot \mathbf{\Tilde{V}}\Vert_\infty \leq \Vert e\Vert_2\sum_{i=1}^d \Vert g_{i,x_i}\Vert_2,
    $$
    where $g_{i,x_i}$ is the $x_i$-derivative of the $i$-th component of $\mathbf{g}$.
\end{lemma}
\begin{proof}
    Expanding $\nabla \cdot \mathbf{\hat{V}}$ into its components (recalling the definition of convolution derivative in Section \ref{sec:math_prel}), and using the Minkowsky inequality (see Lemma \ref{th:Minkowsky}), we can write
    \begin{align}
        \Vert \nabla \cdot \mathbf{\Tilde{V}}\Vert_\infty = \left\Vert \sum_{i=1}^d (g_{i,x_i}*e)\right\Vert_\infty \leq \sum_{i=1}^d  \left\Vert(g_{i,x_i}*e)\right\Vert_\infty,
    \end{align}
    Using Young's convolution inequality, we construct the bound
    \begin{align}
        \Vert \nabla \cdot \mathbf{\Tilde{V}}\Vert_\infty \leq \Vert e\Vert_2\sum_{i=1}^d \Vert g_{i,x_i}\Vert_2,
    \end{align}
    proving the lemma.
\end{proof}
\begin{lemma}\label{lem:lim_sens_lem2}
    If $\nabla \rho^\mathrm{d} \in L^2$, %meaning that the $L^2$-norm of all its components is bounded by a positive constant, 
    i.e. $\Vert \rho^\mathrm{d}_{x_i} \Vert_2 \leq M_i$, for some constants $M_i$ and $i=1,2,3$, then
    $$
    \Vert e\nabla \rho^\mathrm{d} \cdot \mathbf{\hat{V}}\Vert_1 \leq \Vert e\Vert_2^2 \sum_{i=1}^d M_i \Vert g_i\Vert_2,
    $$
    where $g_i$ is the $i$-th component of $\mathbf{g}$.
\end{lemma}
\begin{proof}
    By expanding $\nabla \rho^\mathrm{d}\cdot \mathbf{\hat{V}}$, we get
    \begin{align}\label{eq:first_step}
        \Vert e\nabla \rho^\mathrm{d} \cdot \mathbf{\hat{V}}\Vert_1 = \left\Vert e \sum_{i=1}^d \rho^\mathrm{d}_{x_i} \Tilde{V}_i \right\Vert_1 = \left\Vert e \sum_{i=1}^d \rho^\mathrm{d}_{x_i} (g_i * e)\right\Vert_1.
    \end{align}
    Then, applying Minkowsky (see Lemma \ref{th:Minkowsky}) and the Holder (see Lemma \ref{th:holder}) inequalities, we establish
    \begin{multline}
        \left\Vert e \sum_{i=1}^d \rho^\mathrm{d}_{x_i} (g_i * e)\right\Vert_1 \leq  \sum_{i=1}^d \left\Vert e\rho^\mathrm{d}_{x_i} (g_i * e)\right\Vert_1 \leq\\\leq \sum_{i=1}^d \Vert e\Vert_2\Vert\rho^\mathrm{d}_{x_i}\Vert_2 \Vert(g_i * e)\Vert_\infty.
    \end{multline}
    Finally, applying the Young's covolution inequality, we have
    \begin{align}
        \sum_{i=1}^d \Vert e\Vert_2\Vert\rho^\mathrm{d}_{x_i}\Vert_2 \Vert(g_i * e)\Vert_\infty \leq \sum_{i=1}^d \Vert e\Vert_2^2\Vert\rho^\mathrm{d}_{x_i}\Vert_2 \Vert g_i\Vert_2,
    \end{align}
    which, thanks to the $L^2$-boundedness of $\nabla \rho^\mathrm{d}$ is equivalent to
    \begin{align}\label{eq:fin_step}
        \sum_{i=1}^d \Vert e\Vert_2^2\Vert\rho^\mathrm{d}_{x_i}\Vert_2 \Vert g_i\Vert_2 \leq  \Vert e\Vert_2^2 \sum_{i=1}^d  M_i \Vert g_i\Vert_2 
    \end{align}
    Comparing \eqref{eq:first_step} and \eqref{eq:fin_step} yields the claim. 
\end{proof}

\begin{theorem}[Semiglobal stability with limited sensing]\label{th:LAS_lim_sens}
If $\rho_\mathrm{d}$ and $\nabla \rho_\mathrm{d} \in L^2$, control strategy \eqref{eq:q_lim_sens} achieves semiglobal stabilization of error dynamics \eqref{eq:e_lim_sens}, so that, for any initial condition in the compact set $\Vert e(\cdot, 0)\Vert < \gamma$, choosing $K_\mathrm{p}$ sufficiently large ensures the error to converge asymptotically to 0.
\end{theorem}
\begin{proof}
We choose $\Vert e \Vert_2^2$ as a candidate Lyapunov function. Then, taking into account \eqref{eq:e_lim_sens}, we write (omitting explicit dependencies for simplicity)
\begin{multline}\label{eq:err_lim_sens_dyn}
    (\Vert e \Vert_2^2)_t = \int_\Omega 2ee_t\,\mathrm{d}\mathbf{x} = -2K_\mathrm{p}\Vert e \Vert_2^2+2\int_\Omega e \nabla \cdot (\rho^\mathrm{d}\Tilde{\mathbf{V}})\,\mathrm{d}\mathbf{x}
    \\-2\int_\Omega e \nabla \cdot (e\Tilde{\mathbf{V}})\,\mathrm{d}\mathbf{x}.
\end{multline}
This relation may be rewritten as
\begin{multline}\label{eq:err_lim_sens_2}
    (\Vert e \Vert_2^2)_t = \int_\Omega 2ee_t\,\mathrm{d}\mathbf{x} = -2K_\mathrm{p}\Vert e \Vert_2^2+2\int_\Omega e \nabla \cdot (\rho^\mathrm{d}\Tilde{\mathbf{V}})\,\mathrm{d}\mathbf{x}
    \\-\int_\Omega e^2 \nabla \cdot \Tilde{\mathbf{V}}\,\mathrm{d}\mathbf{x},
\end{multline}
where, applying Lemma \ref{lem:div_rel}, the divergence theorem, and Lemma \ref{lem:surf_int}, we establish  
\begin{multline}\label{eq:err_norm_lim_sens_01}
    2\int_\Omega e \nabla \cdot (e\Tilde{\mathbf{V}})\,\mathrm{d}\mathbf{x} = 2 \int_\Omega \nabla  \cdot (e^2\Tilde{\mathbf{V}})\,\mathrm{d}\mathbf{x}-2\int_\Omega \nabla e \cdot (e \Tilde{\mathbf{V}})\,\mathrm{d}\mathbf{x}\\=2 \int_{\partial\Omega} e^2 \Tilde{\mathbf{V}}\cdot \mathbf{\hat{n}}\,\mathrm{d}\mathbf{x}-2\int_\Omega \nabla e \cdot (e \Tilde{\mathbf{V}})\,\mathrm{d}\mathbf{x}=-2 \int_\Omega \nabla e \cdot (e\Tilde{\mathbf{V}})\,\mathrm{d}\mathbf{x} \\= -\int_\Omega \nabla(e^2)\cdot\Tilde{\mathbf{V}}\,\mathrm{d}\mathbf{x}=-\int_\Omega \nabla \cdot (e^2\Tilde{\mathbf{V}})\,\mathrm{d}\mathbf{x} + \int_\Omega e^2 \nabla \cdot \Tilde{\mathbf{V}}\,\mathrm{d}\mathbf{x}\\= -\int_{\partial\Omega}e^2\Tilde{\mathbf{V}}\cdot\mathbf{\hat{n}}\,\mathrm{d}\mathbf{x}+ \int_\Omega e^2 \nabla \cdot \Tilde{\mathbf{V}}\,\mathrm{d}\mathbf{x} =\int_\Omega e^2 \nabla \cdot \Tilde{\mathbf{V}}\,\mathrm{d}\mathbf{x}.
\end{multline}
% We remark that, by means of the Young's convolution inequality and Minkowsky inequality, we can establish
% \begin{multline}
%     \Vert \nabla \cdot \mathbf{\Tilde{V}} \Vert_\infty = \Vert(g_{1x} * e)+(g_{2y} * e)+(g_{3z} * e)\Vert_\infty \leq\\\leq \Vert e \Vert_2 \Vert g_{1x} +g_{2y}+g_{3z}\Vert_2,
% \end{multline}
% where $g_1$, $g_2$, and $g_3$ are the components of $\mathbf{g}$. 
%Consequently, we know that 
% \begin{multline}
%     \Vert \nabla \cdot (\rho^\mathrm{d}\mathbf{\Tilde{V}} )\Vert_\infty = \Vert \rho^\mathrm{d}\nabla \cdot \mathbf{\Tilde{V}} + \nabla\rho^\mathrm{d} \cdot \mathbf{\Tilde{V}}  \Vert_\infty \leq L
% \end{multline}
We can provide bounds for the last two terms of \eqref{eq:err_lim_sens_2}, namely
\begin{multline}\label{eq:lim_sense_bound1}
    \left\vert \int_\Omega e \nabla \cdot (\rho^\mathrm{d}\Tilde{\mathbf{V}})\,\mathrm{d}\mathbf{x} \right\vert \leq  \int_\Omega \left\vert e \nabla \cdot (\rho^\mathrm{d}\Tilde{\mathbf{V}})\right\vert\,\mathrm{d}\mathbf{x} \\= \Vert e \nabla \cdot (\rho^\mathrm{d}\Tilde{\mathbf{V}})\Vert_1 = \Vert e\rho^\mathrm{d} \nabla \cdot \mathbf{\Tilde{V}} + e \nabla\rho^\mathrm{d}\cdot \mathbf{\Tilde{V}}\Vert_1 \leq\\\leq \Vert e\rho^\mathrm{d} \nabla \cdot \mathbf{\Tilde{V}}\Vert_1 +\Vert e \nabla\rho^\mathrm{d}\cdot \mathbf{\Tilde{V}}\Vert_1 \leq \Vert e \Vert_2 \Vert\rho^\mathrm{d} \Vert_2\Vert \nabla \cdot \mathbf{\Tilde{V}} \Vert_\infty \\+ \Vert e\Vert_2^2 \sum_{i=1}^d  M_i \Vert g_i\Vert_2 
    \leq \Vert e\Vert_2^2 \left(  \sum_{i=1}^d L\Vert g_{i,x_i}\Vert_2 + M_i \Vert g_i\Vert_2  \right),
    % \leq LG \Vert e\Vert_2^2 + \Vert e\Vert_2 \Vert \rho_x^\mathrm{d}\Vert_2 \Vert(g_1*e) +  \Vert e\Vert_2 \Vert \rho_y^\mathrm{d}\Vert_2 \Vert(g_2*e) \\+ \Vert e\Vert_2 \Vert \rho_z^\mathrm{d}\Vert_2 \Vert(g_3*e)  \Vert_\infty\leq\\
    % \leq LG \Vert e\Vert_2^2 + (M_1 \Vert g_1\Vert_2 + M_2\Vert g_2\Vert_2 + M_3\Vert g_3\Vert_2) \Vert e\Vert_2^2,
\end{multline}
\begin{multline}\label{eq:lim_sense_bound2}
    \left\vert \int_\Omega e^2 \nabla \cdot \Tilde{\mathbf{V}}\,\mathrm{d}\mathbf{x} \right\vert \leq  \int_\Omega \left\vert e^2 \nabla \cdot \Tilde{\mathbf{V}}\right\vert\,\mathrm{d}\mathbf{x} = \Vert e^2 \nabla \cdot \Tilde{\mathbf{V}}\Vert_1 \leq\\\leq \Vert e \Vert_2^2 \Vert \nabla \cdot \Tilde{\mathbf{V}}\Vert_\infty  \leq\Vert e \Vert_2^3 \sum_{i=1}^d \Vert g_{i,x_i}\Vert_2,
\end{multline}
where $L$ is a positive constant bounding $\Vert\rho^\mathrm{d}\Vert_2$, and we used Lemma \ref{lem:lim_sens_lem1} and \ref{lem:lim_sens_lem2}, as well as the Holder's inequality. 
Ultimately, we establish that
\begin{align}
    (\Vert e \Vert_2^2)_t \leq (-2K_\mathrm{p} + F + G\Vert e \Vert_2)\Vert e \Vert_2^2, 
\end{align}
where 
\begin{align}
    F &= 2 \sum_{i=1}^d L\Vert g_{i,x_i}\Vert_2 + M_i \Vert g_i\Vert_2,\\
    G &= \sum_{i=1}^d \Vert g_{i,x_i}\Vert_2.
\end{align}
Choosing $K_\mathrm{p}>(F+G\gamma)/2$, the error asymptotically converges to 0. 
\end{proof}

\section{Structural Perturbations}\label{sec:pert}
\begin{figure*}
\centering
\subfloat[]{\includegraphics[width=0.25\textwidth]{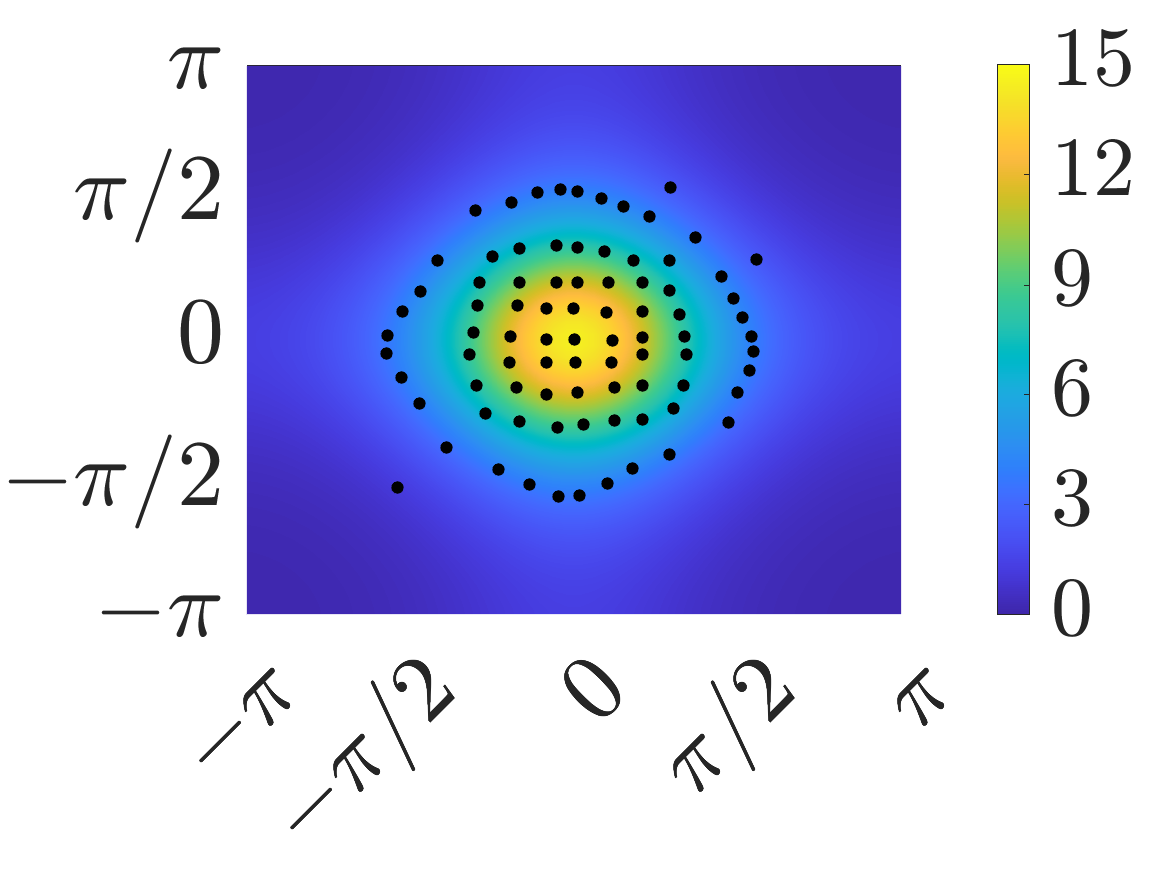}%
\label{subfig:fin_disp_lim}}
%\hspace{0.2cm}
\subfloat[]{\includegraphics[width=0.25\textwidth]{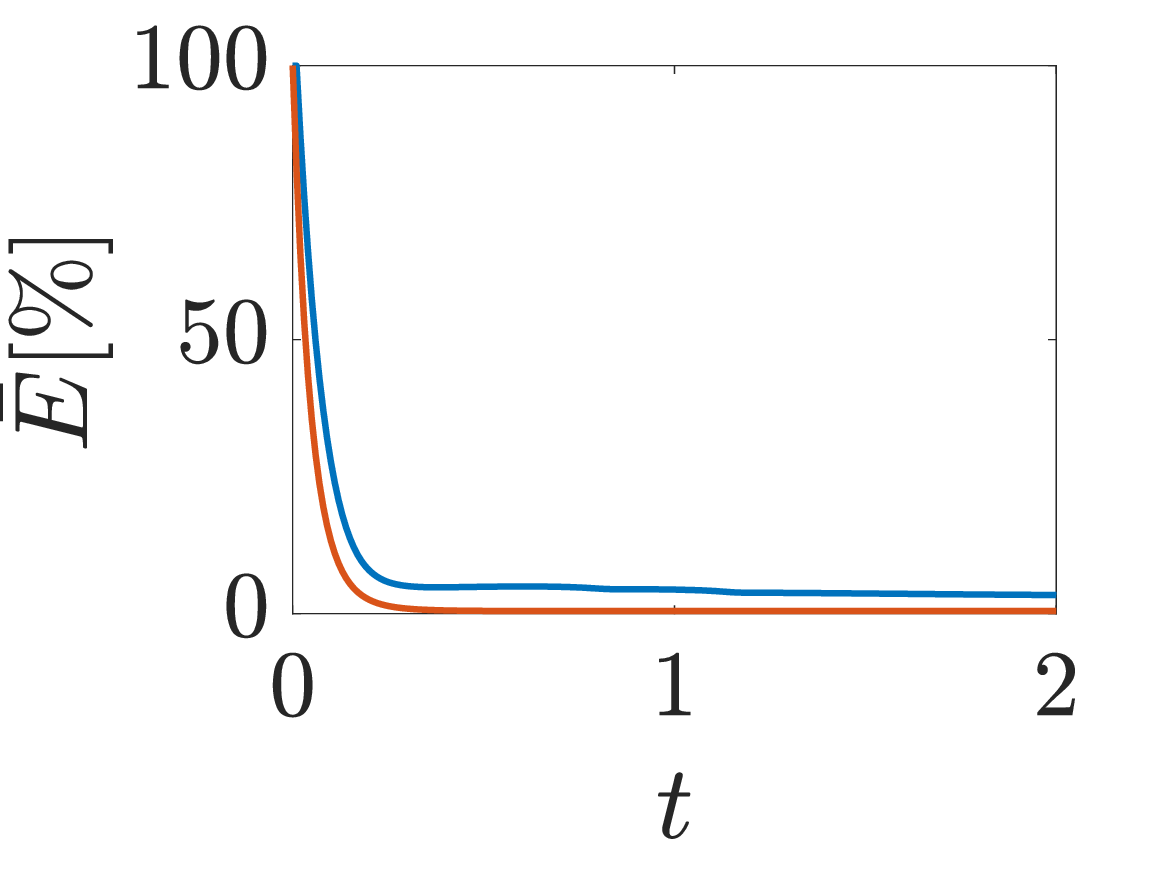}%
\label{subfig:dkl_fin_disp}}
%\hspace{0.2cm}
\subfloat[]{\includegraphics[width=0.25\textwidth]{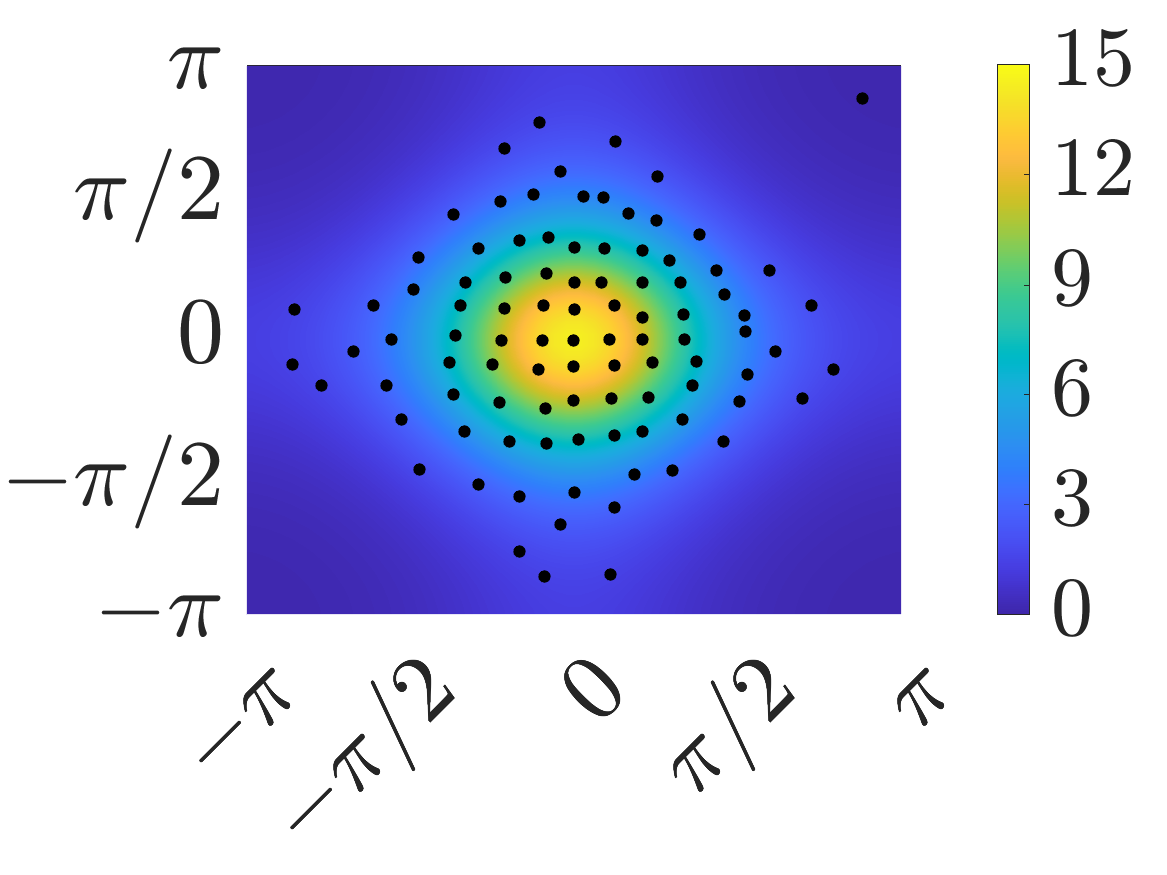}%
\label{subfig:findisp_unlim}}
%\hspace{0.2cm}
\subfloat[]{\includegraphics[width=0.25\textwidth]{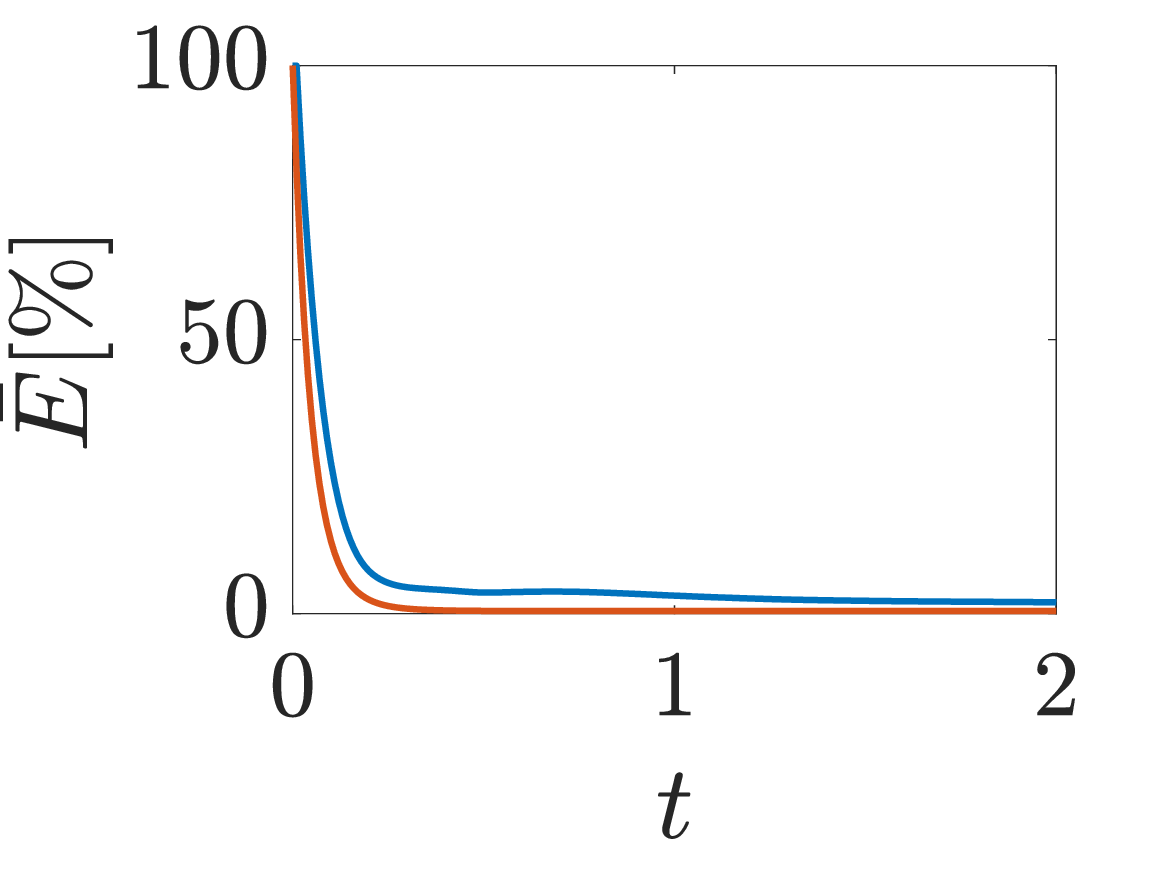}%
\label{subfig:dkl_unlim}}
\caption{Robustness to limited sensing ($K_\mathrm{p} = 100$, (a, b) $\Delta = 0.1\pi$, (c, d) $\Delta = \pi$). (a, c) Final displacement of the system on top of the desired density; (b, d) time evolution of the percentage error for the discrete (blue) and continuous case (orange).}
\label{fig:lim_sens}
\end{figure*}
Here, we assume our system to be perturbed by spatio-temporal disturbances. In particular, we study
\begin{align}
    \rho_t(\mathbf{x}, t) + \nabla \cdot \left[ \rho(\mathbf{x}, t)\left(\mathbf{V}(\mathbf{x}, t)+\mathbf{W}(\mathbf{x}, t)\right)\right] = q(\mathbf{x}, t),
\end{align}
where $\mathbf{W}$ is a perturbing velocity field. Further, we hypothesize ($i$) $\mathbf{W}$ to be periodic on $\partial \Omega$, ($ii$) components of $\mathbf{W}$ to be $L^\infty$ bounded by some positive constants $\bar{W}_i$ ($i=1,2,3$), and ($iii$)
$\Vert\nabla \cdot \mathbf{W}\Vert_\infty\leq\widehat{W}$ .
In such a scenario, the error dynamics become
\begin{multline}
    e_t(\mathbf{x}, t) = -K_\mathrm{p}e(\mathbf{x}, t) + \nabla \cdot [\rho^\mathrm{d}(\mathbf{x}, t) \mathbf{W}(\mathbf{x}, t)]\\- \nabla \cdot [e(\mathbf{x}, t) \mathbf{W}(\mathbf{x}, t)].
\end{multline}

\begin{theorem}[Bounded stability with perturbations]\label{th:perturbations}
    $\,$In the presence of a bounded spatio-temporal disturbance $\mathbf{W}$, and if $\Vert \rho^\mathrm{d}\Vert_2 \leq L$ and $\Vert \rho^\mathrm{d}_{x_i}\Vert \leq M_i$  ($i=1, \dots, d$), there exists a threshold value $\kappa > 0$, such that for $K_\mathrm{p}>\kappa$, the dynamics of $\Vert e\Vert_2^2$ remains bounded. Specifically,
    \begin{align}\label{eq:err_bound_pert}
        \lim_{t\to\infty} \sup \Vert e(\cdot, t)\Vert_2 \leq \frac{H}{2K_\mathrm{p} - \widehat{W}},
    \end{align}
    with $H = 2 \left(L\widehat{W} + \sum_{i=1}^d M_i\bar{W}_i\right)$.
\end{theorem}
\begin{proof}
    We write the dynamics of $\Vert e \Vert_2^2$ as
    \begin{multline}\label{eq:norm_err_dist_0}
        (\Vert e \Vert_2^2)_t = 2\int_\Omega ee_t\,\mathrm{d}\mathbf{x} = -2K_\mathrm{p} \Vert e\Vert_2^2 + 2 \int_\Omega e \nabla \cdot (\rho^\mathrm{d}\mathbf{W})\,\mathrm{d}\mathbf{x} \\- 2 \int_\Omega e \nabla \cdot (e\mathbf{W})\,\mathrm{d}\mathbf{x}.
    \end{multline}
    Similarly to the proof of Theorem \ref{th:LAS_lim_sens} (specifically \eqref{eq:err_lim_sens_2} and \eqref{eq:err_norm_lim_sens_01}), we can rewrite \eqref{eq:norm_err_dist_0} as
    \begin{multline}
        (\Vert e\Vert_2^2)_t = 2\int_\Omega ee_t\,\mathrm{d}\mathbf{x} = -2K_\mathrm{p} \Vert e\Vert_2^2 + 2 \int_\Omega e \nabla \cdot (\rho^\mathrm{d}\mathbf{W})\,\mathrm{d}\mathbf{x} \\- \int_\Omega e^2\nabla \cdot \mathbf{W}\,\mathrm{d}\mathbf{x}.
    \end{multline}
    Similarly to \eqref{eq:lim_sense_bound1} and \eqref{eq:lim_sense_bound2} in the proof of Theorem \ref{th:LAS_lim_sens}, we can give the bounds
    \begin{multline}
        \left\vert \int_\Omega e \nabla \cdot (\rho^\mathrm{d}\mathbf{W})\,\mathrm{d}\mathbf{x}\right\vert \leq \int_\Omega \left\vert e \nabla \cdot (\rho^\mathrm{d}\mathbf{W})\right\vert\,\mathrm{d}\mathbf{x} \\= \Vert e \nabla \cdot (\rho^\mathrm{d}\mathbf{W})\Vert_1 = \Vert e\rho^\mathrm{d} \nabla\cdot\mathbf{W} + e\nabla\rho^\mathrm{d}\cdot\mathbf{W}\Vert_1 \leq\\
        \leq\Vert e\rho^\mathrm{d} \nabla\cdot\mathbf{W}\Vert_1 + \Vert e\nabla\rho^\mathrm{d}\cdot\mathbf{W}\Vert_1 \leq \Vert e\Vert_2\Vert \rho^\mathrm{d}\Vert_2 \Vert\nabla\cdot\mathbf{W} \Vert_\infty+\\+\Vert e \Vert_2 \sum_{i=1}^d \Vert \rho_{x_i}^\mathrm{d}\Vert_2\Vert W_i \Vert_\infty \leq \frac{H}{2}\Vert e\Vert_2,
    \end{multline}
    \begin{multline}
        \left\vert \int_\Omega e^2\nabla \cdot \mathbf{W}\,\mathrm{d}\mathbf{x}\right\vert \leq \int_\Omega \left\vert e^2\nabla \cdot \mathbf{W}\,\mathrm{d}\mathbf{x}\right\vert = \left\Vert e^2\nabla \cdot \mathbf{W}\right\Vert_1\leq\\\leq \Vert e \Vert_2\Vert e \Vert_2\Vert \nabla \cdot \mathbf{W}\Vert_\infty\leq\widehat{W}\Vert e \Vert_2^2 
    \end{multline}
    % where
    % \begin{align}
    %     H= 2 \left(L\widehat{W} + \sum_{i=1}^d M_i\bar{W}_i\right),
    % \end{align}
    % $L$, as in Theorem \ref{th:LAS_lim_sens} is a positive constant bounding $\Vert \rho^\mathrm{d}\Vert_2$, and $M_i$ is a positive constant bounding $\Vert \rho^\mathrm{d}_{x_i}\Vert_2$.
    Then, setting $\eta = \Vert e \Vert_2^2$, we establish
    \begin{align}
        \eta_t \leq -A \eta + H\sqrt{\eta},
    \end{align}
    where $H$ is given in the theorem statement, and $A = 2K_\mathrm{p} - \widehat{W}$. If we assume $A$ to be positive, i.e., $2K_\mathrm{p}>\widehat{W}$, the bounding field is exhibiting a global asymptotically stable equilibrium point at $H^2/A^2$. Then, thanks to the Lemma \ref{lemma:comparison_lemma}, \eqref{eq:err_bound_pert} is recovered.
    % \begin{align}
    %     \lim_{t\to\infty} \sup \Vert e(\cdot, t)\Vert_2 \leq \frac{H}{A} = \frac{H}{2K_\mathrm{p} - \widehat{W}}.
    % \end{align}
    Hence, if $K_\mathrm{p}>\kappa>\widehat{W}/2$, $\Vert e\Vert_2$ remains bounded by $H/A$.
\end{proof}

% \textbf{Numerical validation:} using the same numerical set-up that is discussed in Section \ref{sec:lim_sens}, we consider here a step disturbance of amplitude $\hat{d}$ on both the $x$ and $y$ direction coming at half of the trial, that is $\mathbf{W}(x, t) = \hat{d} \,[\mathrm{step}(t-t\mathrm{f}/2), \mathrm{step}(t-t\mathrm{f}/2)]^T$.\\
% \indent In this case, for a trial of 400 time steps, we observe the results in Fig. \ref{subfig:dist_cont} for the continuous case and Fig. \ref{subfig:dist_disc} for the discrete one. In both scenarios, we observe that, when the perturbation kicks in, the KL divergence settles to a bounded value, confirming what is shown in Theorem \ref{th:perturbations}. 

\section{Numerical validation}\label{sec:num_valid}
\begin{figure}
\centering
\subfloat[]{\includegraphics[width=0.4\textwidth]{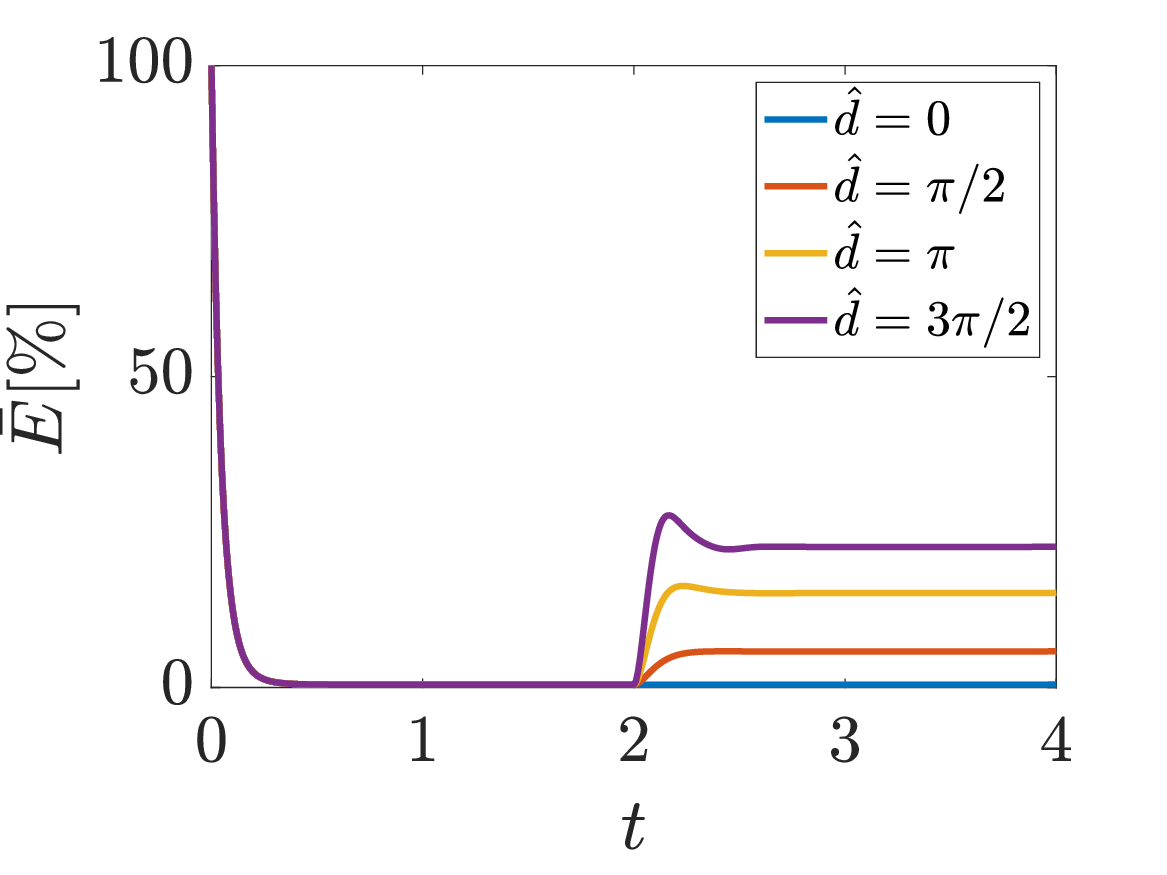}%
\label{subfig:dist_cont}}
%\hspace{0.2cm}

\subfloat[]{\includegraphics[width=0.4\textwidth]{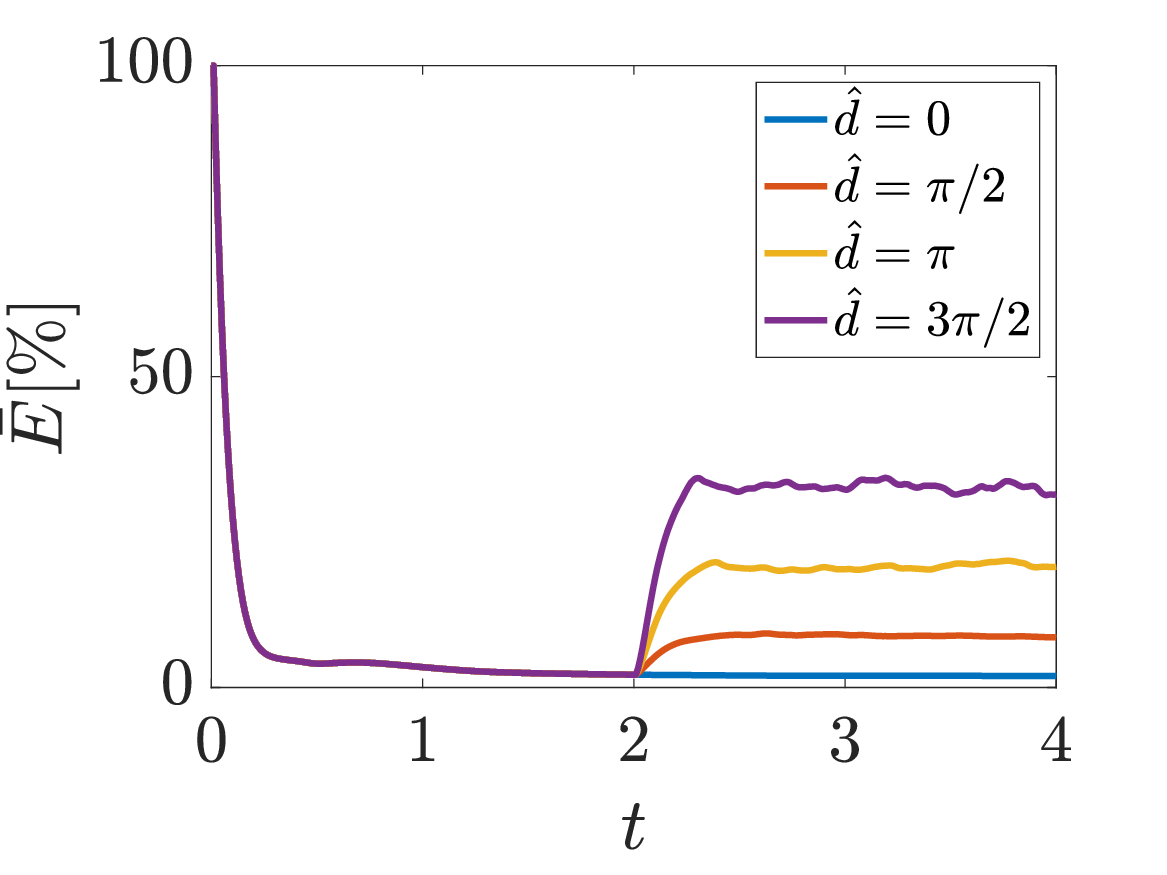}%
\label{subfig:dist_disc}}
%\hspace{0.2cm}
\caption{Robustness to perturbations ($K_\mathrm{p} = 100$): percentage error in time for (a) a continuous trial, (b) a discrete trial.}
\label{fig:lim_disturbances}
\end{figure}
For numerical validation, we adopt the  configuration presented in \cite{carlo2023mixed}. In this setup, in the absence of control mechanisms, agents engage through a repulsive periodic kernel, leading to their dispersal across the domain towards a uniform density. The target density configuration to be achieved is a 2D von Mises distribution with its center at the domain's origin. For details on this setup, the reader is directed to \cite{carlo2023mixed}.

We consider a sample of 100 agents starting from a constant density profile, and, for each trial, we run both a discrete and a continuous simulation. This means that, in every scenario, we numerically integrate both \eqref{eq:d_dimensional_micro} and its continuified version \eqref{eq:d_dimensional_macro}, allowing us to understand how well the continuum approximation holds. For the discrete trials, we use forward Euler with $\Delta t = 0.001$, and, for the computation of spatial functions involved in the definition of $\mathbf{u}_i$, we discretize $\Omega$ using a mesh of 50$\times$50 cells. We remark that agents are not constrained to move on this mesh, that is merely used for evaluating spatial functions. For the numerical integration of the continuified problem, we use a Lax-Friedrichs finite volumes scheme \cite{leveque2002finite} with the same $\Delta t$ and spatial mesh of the discrete simulations.

The performance of the trials is assessed using the normalized percentage error
\begin{align}
    \bar{E}(t) = \frac{\Vert e (\cdot, t)\Vert_2^2}{\max_t \Vert e (\cdot, t)\Vert_2^2}100.
\end{align}

\paragraph{Robustness to limited sensing} to validate the stability result of Theorem \ref{th:LAS_lim_sens}, we fix $K_\mathrm{p} = 100$. When running a trial of 200 time steps, we obtain the results in Fig. \ref{subfig:fin_disp_lim}, \ref{subfig:dkl_fin_disp} choosing $\Delta = 0.1 \pi$ (i.e., agents have a sensing radius of 10\% of the domain), and those in Fig. \ref{subfig:findisp_unlim}, \ref{subfig:dkl_unlim} with $\Delta = \pi$ (i.e., unlimited sensing). This choice of $K_\mathrm{p}$ ensures the performance is independent of the sensing capabilities of the agents. In the discrete trials, we observe a non-zero steady-state error. This is due to the finite-size effect of assuming a swarm of 100 agents. This residual error is slightly worse in the case of limited sensing.

\paragraph{Robustness to perturbations} to numerically assess robustness to perturbations, we consider a step disturbance of amplitude $\hat{d}$ on both the $x$ and $y$ direction coming at half of the trial, that is $\mathbf{W}(x, t) = \hat{d} \,[\mathrm{step}(t-t\mathrm{f}/2), \mathrm{step}(t-t\mathrm{f}/2)]^T$.
In this case, for a trial of 400 time steps, we observe the results in Fig. \ref{subfig:dist_cont} for the continuous case and Fig. \ref{subfig:dist_disc} for the discrete one. For both scenarios, we observe that, when the perturbation is active, the error settles to a bounded value, confirming findings in Theorem \ref{th:perturbations}. We also remark that the error settles well below the theoretical estimate of Theorem \ref{th:perturbations}, for example, when $\hat{d} = 3\pi/2$, $H/A \approx 0.8$, while $\Vert e(\cdot, t_\mathrm{f})\Vert_2 \approx 0.1$.  

\section{Conclusions}
Upon recalling the theoretical framework presented in \cite{carlo2023mixed}, we analytically assessed the robustness of the control solution with respect to limited sensing capabilities and perturbations. We demonstrated that, out of the nominal condition, stability can still be preserved.

\bibliographystyle{IEEEtran}

\end{document}